\documentclass{article}
\usepackage{spconf,amsmath,graphicx}

\usepackage{fancyhdr}

\fancypagestyle{firstpagewide}{
  \fancyhf{}
  \fancyfoot[C]{\parbox{\textwidth}{\centering\scriptsize
  This work has been submitted to the IEEE for possible publication. Copyright may be transferred without notice, after which this version may no longer be accessible.}}

}

\usepackage{booktabs}
\usepackage{makecell}
\usepackage{tabularx}
\usepackage{array}
\usepackage{multirow}
\usepackage{hyperref}
\usepackage{hyphenat}
\usepackage{xcolor}
\usepackage{soul}  
\usepackage{refcount} 

\usepackage{enumitem}
\setlist{nosep,leftmargin=14pt}


\usepackage{afterpage}
\usepackage{pdfpages}
\usepackage[section]{placeins}
\usepackage{dblfloatfix} 

\graphicspath{{Paper\space Material/figures/}}

\title{Computational Imaging Meets LLMs: Zero-Shot IDH Mutation Prediction in Brain Gliomas}

\name{Syed Muqeem Mahmood, Hassan Mohy-ud-Din}
\address{School of Science and Engineering, Lahore University of Management Sciences, Lahore, Pakistan}

\begin{document}
\ninept
\maketitle
\begin{abstract}
We present a framework that combines Large Language Models with computational image analytics for non-invasive, zero-shot prediction of IDH mutation status in brain gliomas. For each subject, co-registered multi-parametric MRI scans and multi-class tumor segmentation maps were processed to extract interpretable semantic (visual) attributes and quantitative features, serialized in a standardized JSON file, and used to query GPT-4o and GPT-5 without fine-tuning. We evaluated this framework on six publicly available datasets (\(N=1427\)) and results showcased high accuracy and balanced classification performance across heterogeneous cohorts – even in the absence of manual annotations. GPT-5 outperformed GPT-4o in context-driven phenotype interpretation. Volumetric features emerged as the most important predictors, supplemented by subtype-specific imaging markers and clinical information. Our results demonstrate the potential of integrating LLM-based reasoning with computational image analytics for precise, non-invasive tumor genotyping, advancing diagnostic strategies in neuro-oncology.
The code is available at \texttt{https://github.com/ATPLab-LUMS/CIM-LLM}.
\end{abstract}

\section{Introduction}
Gliomas are the most common brain neoplasms that originate in the glial cells of the brain. They account for \(30\)\% of all brain and CNS neoplasms and \(80\)\% of malignant brain tumors The World Health Organization (WHO) classifies them into four grades based on their aggressiveness \cite{louis20212021}: WHO Grade II and Grade III are referred to as Low-grade Gliomas (LGGs) and WHO Grade IV represents High-grade Gliomas (HGGs). According to WHO criteria, the prognosis and treatment strategies of brain gliomas are strongly influenced by molecular markers, particularly the isocitrate dehydrogenase (IDH) mutation status \cite{louis20212021}. Distinguishing IDH-mutant from IDH-wildtype gliomas is therefore crucial for accurate diagnosis, personalized treatment planning, and reliable prognostication.

The gold standard approach for obtaining IDH mutation status is invasive biopsy or resection, followed by expensive molecular testing and analysis. Biopsy or resection is not always feasible due to tumor location, patient condition, or resource constraints. This presses the need for non-invasive prediction of IDH mutation status from preoperative MRI scans, which could greatly enhance clinical decision-making, particularly in underprivileged, remote, or resource-limited healthcare settings \cite{rudie2019emerging}.

3D multiparametric MRI scans (3D mpMRI) captures rich phenotypic information about tumor morphology, enhancement, and location. Prior studies have identified radiologic correlates of IDH mutation, including non-enhancing components, regular shape, and the T2–FLAIR mismatch sign \cite{su2020radiomics, chow2018imaging, smits2017imaging}. Yet, traditional machine learning or radiomics models rely on large labeled datasets and task-specific retraining, limiting generalization across diverse imaging protocols and institutions and reduces interpretability \cite{lohmann2022radiomics}. A generalizable, zero-shot approach capable of interpreting imaging-derived features without retraining could overcome these challenges.

Large Language Models (LLMs) such as GPT-4o and GPT-5 demonstrate strong reasoning and contextual understanding across modalities. By integrating semantic (visual) attributes, image-derived quantitative features, and clinical information, LLMs may perform high-level interpretation tasks typically reserved for expert radiologists \cite{kang2024mri}. However, their potential for structured medical imaging analysis, particularly genotype prediction, remains underexplored.

This study presents a framework which combines LLMs-based reasoning with computational image analytics for zero-shot IDH genotype prediction in brain gliomas. For each subject, co-registered 3D mpMRI volumes and multi-class tumor segmentation maps were processed to extract interpretable semantic (visual) and quantitative features, serialized in a standardized JSON file, and used to query GPT-4o and GPT-5 without fine-tuning. We evaluated this framework on six publicly available datasets and results showcased high accuracy and balanced classification performance across heterogeneous cohorts – even in the absence of manual annotations – highlighting enormous potential of LLMs to integrate multimodal imaging and clinical contextual information for non-invasive, zero-shot molecular characterization.

\section{Methodology}
\label{sec:method}

\begin{figure}[t]
  \centering
  \includegraphics[width=\linewidth]{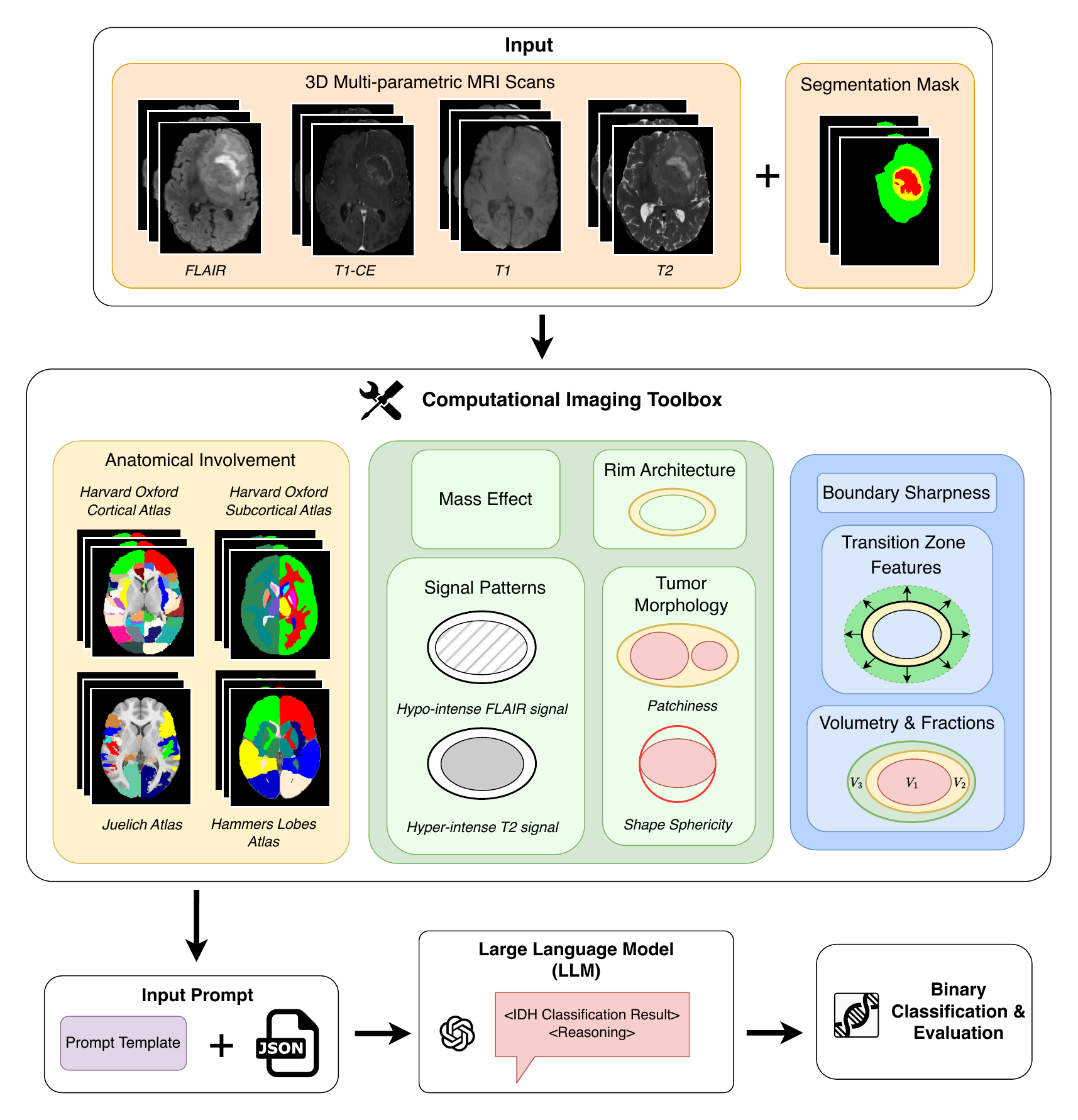}
  \caption{Flowchart of the proposed framework.}
  \label{fig:main_flowchart}
\end{figure}

Figure~\ref{fig:main_flowchart} illustrates the proposed framework. For each subject, co-registered 3D mpMRI volumes and corresponding multiclass segmentation maps are processed to extract a set of radiologically interpretable visual and quantitative attributes. These attributes are serialized following a predefined JSON schema and combined with a task-specific prompt to query a LLM. The LLM’s response is then parsed to predict the IDH mutation status (mutant vs. wild type).

\subsection{Multimodal Imaging Data, Annotations, and Preprocessing}
For each subject, we used 3D mpMRI volumes (FLAIR, T1-weighted, contrast-enhanced T1-weighted, and T2-weighted sequences) along with a co-registered multiclass tumor segmentation map following the BraTS 2021 annotation protocol \cite{baid2021rsna}. For cases without manual labels, segmentations were automatically generated using the BrainSegFounder foundation model \cite{cox2024brainsegfounder}, with per-voxel class probabilities aggregated across all available cross-validation folds by soft voting. All MRI sequences and masks were rigidly registered (6-DOF) to the MNI-152 template \cite{fonov2009unbiased, avants2011reproducible}, resampled to \(1\times1\times1\ \mathrm{mm}^3\), and corrected for bias-field nonuniformity using the N4 algorithm. Preprocessing was performed with ANTsPy\footnote{\href{https://antspy.readthedocs.io/en/stable/}{ANTsPy documentation}}.

\subsection{Discriminating Features: IDH-mutant vs IDH-wildtype}
IDH-mutant gliomas typically exhibit a predominantly non-enhancing component with minimal or no contrast uptake, a more regular (spherical) morphology, and sharper margins with a thin transition zone \cite{smits2017imaging, azizova2024ten}. When present, the T2--FLAIR mismatch sign is highly specific for this subtype \cite{jain2020real}. These tumors also tend to have less edema and appear more often in the frontal lobe with limited contact to the subventricular zone \cite{azizova2024ten, park2018prediction}.

In contrast, IDH--wildtype gliomas are more centrally located with frequent subventricular zone contact, larger enhancing and necrotic portions relative to non-enhancing components, irregular morphology with low sphericity, and poorly defined margins with a thick transition zone \cite{smits2017imaging, azizova2024ten, park2018prediction}. They also more often cross the midline and show a greater edema burden with associated ventricular compression \cite{palmisciano2022gliomas, azizova2024ten}.

\subsection{Computational Imaging Toolbox}
\label{sec:toolbox}
We developed a modular pipeline in Python to extract distinguishing semantic (visual) attributes and quantitative features from co-registered 3D mpMRI volumes and corresponding multiclass segmentation masks. All measures were computed in the MNI space, expressed in physical units (\(\mathrm{mm}, \mathrm{mL}\)), and normalized where appropriate. The acquired set of measures are broadly classified into five categories described below.

\paragraph*{Location Features.}
Tumor localization was analyzed using four MNI-152 atlases: Harvard–Oxford cortical/subcortical \cite{nilearn_harvard_oxford}, Jülich cytoarchitectonic \cite{nilearn_juelich}, and Hammers lobar \cite{dadar2023hammers}. Atlas maps were resampled to MNI space and non-rigidly registered with ANTsPy (Affine + SyN). For each region, tumor-in-region and regional occupancy were computed from the tumor core (TC). Eloquent-cortex proximity was quantified as the Euclidean distance from the TC boundary to each eloquent region, reporting the five nearest neighbors. Deep gray nuclei and bilateral frontal involvement were encoded as binary variables, and the edema subregion was analyzed separately.

\paragraph*{T2-FLAIR Mismatch}
From bias-corrected, co-registered T2-weighted and FLAIR images, three features were extracted per subject: (1) \textit{FLAIR signal suppression} (binary)---a homogeneous, T2-hyperintense, non-enhancing tumor (NET) subregion showing relative FLAIR hypointensity; (2) \textit{FLAIR rim hyperintensity} (binary)---a FLAIR-hyperintense rim along the enhancing tumor (ET) margin, relative to the suppressed NET; and (3) \textit{T2-FLAIR mismatch ratio}---the ratio of normalized T2 to FLAIR intensities within the NET. Intensity normalization used the median signal of contralateral normal-appearing white matter (CNWM), automatically segmented via the Deep Atropos model in ANTsPyNet\footnote{\href{https://antsx.github.io/ANTsPyNet/}{ANTsPyNet documentation}\label{fn:antspynet}}.

\paragraph*{Mass Effect}
Midline crossing was reported for the TC and edema (ED) subregions as binary variables. Ventricular compression was assessed from T1-weighted images by segmenting left and right lateral ventricles with the ANTsPyNet DKT model\footnotemark[\getrefnumber{fn:antspynet}]. Ventricular volumes derived from these binary masks were used to compute an asymmetry index, where magnitude reflected compression severity/asymmetry and sign indicated lateralization (right \(>\) left for positive values).

\paragraph*{Tumor Morphology.}
Compact shape and organization descriptors were derived in physical units based on voxel spacing (\(\mathrm{mm}\)): (1) \textit{Rim architecture} included measures of TC hollowness, rim–core adjacency (fraction of core boundary voxels in contact with rim), and enhancing rim thickness (\(\mathrm{mm}\)); (2) \textit{Patchiness} captured the number of disconnected components within the ET and NET subregions, as well as the non-rim enhancement fraction (proportion of ET voxels not adjacent to NET); (3) \textit{Morphology} included sphericity of the whole tumor (WT) and TC, and boundary sharpness indices \cite{Olafson2021BSC} for WT and TC derived from local intensity gradients; and (4) \textit{Transition Zone Thickness} quantified the spatial extent (in \(\mathrm{mm}\)) over which TC intensity gradually transitions to the characteristic intensity of the surrounding ED subregion \cite{Price2006DTI}. For robustness, median values were employed where appropriate.

\paragraph*{Volumetric Measures.}
Volumes were measured for the WT, NET, ET, and ED subregions. Fractional burden metrics described subregional composition, including NET and ET proportions within WT, enhancement within the TC, and edema burden relative to NET and TC. Edema spatial extent was summarized as the distance (median and \(95\)\(^{\mathrm{th}}\) percentile) from the TC boundary to surrounding ED voxels, computed in physical space using voxel spacing.

\paragraph*{}
All semantic (visual) attributes and quantitative measures were serialized in a JSON file. Missing attributes/features were reported as \texttt{NULL}. We also added clinical features (gender and age) to the JSON file.

\subsection{Large Language Models for Zero-shot Inference}
\label{sec:prompt_construct}
We evaluated two LLMs: GPT-4o (\texttt{gpt-4o}) \cite{openai_gpt4o_modelcard} and GPT-5 Chat (\texttt{gpt-5-chat-latest}) \cite{openai_gpt5_chat_latest_modelcard} for zero-shot prediction of IDH mutation status using the subject-specific JSON file as input. The following prompt was used:

\textit{You are an experienced radiologist entasked with discriminating a brain glioma as either `IDH mutant' or `IDH wildtype'. You are presented a JSON file encapsulating semantic (visual) attributes and quantitative metrics about a brain tumor (glioma)---extracted from 3D multiparametric MRI sequences (FLAIR, T1-contrast enhanced, and T2-weighted) and a co-registered 3D segmentation map of tumor subregions. Note that we do not have information on the necrosis component of the tumor. Provide a compact response with compact reasoning. Structure your response as follows:  \textless **Final IDH type** \textgreater \textbackslash n  \textless Reasoning\textgreater.}

For binary classification, the model was constrained to output a boldfaced class label on the first line followed by a brief rationale. The predicted label was extracted using regular expressions, mapping normalized variants (e.g., “IDH mutant”, “mutated”, “IDH mutation” \(\rightarrow\) 1; “IDH wildtype”, “wildtype” \(\rightarrow\) 0).

\newcolumntype{R}{>{\raggedleft\arraybackslash}X}
\begin{table}[t]
\centering
\footnotesize
\setlength{\tabcolsep}{3pt}
\renewcommand{\arraystretch}{1.15}
\newcolumntype{C}{>{\centering\arraybackslash}X}
\caption{Patient characteristics}
\label{tab:overall_characteristics}
\begin{tabularx}{0.8\linewidth}{@{}l C@{}}
\toprule
\textbf{Characteristic} & \textbf{Overall cohort (\textit{n} = 1427)} \\
\midrule
\multicolumn{2}{@{}l}{\textbf{Age (years)}}\\
Known   & \(58.2 \pm 14.47\) \\
Unknown & 6 (0.4\%) \\
\midrule
\multicolumn{2}{@{}l}{\textbf{Sex}}\\
Female & 579 (41\%) \\
Male   & 848 (59\%) \\
\midrule
\multicolumn{2}{@{}l}{\textbf{IDH}}\\
Mutated  & 295 (21\%) \\
Wildtype & 1132 (79\%) \\
\midrule
\multicolumn{2}{@{}l}{\textbf{Molecular Subtype}}\\
Oligodendroglioma         & 109 (8\%) \\
IDH-mutant astrocytoma    & 186 (13\%) \\
IDH-wildtype glioblastoma & 1132 (79\%) \\
\midrule
\multicolumn{2}{@{}l}{\textbf{Tumor Grade}}\\
WHO grade 2 & 198 (14\%) \\
WHO grade 3 & 97  (7\%) \\
WHO grade 4 & 1132 (79\%) \\
\bottomrule
\end{tabularx}
\end{table}

\section{Experiments}
\newcolumntype{Y}{>{\centering\arraybackslash}X}
\begin{table*}[!t]
  \centering
  \caption{Performance across datasets (95\% CIs in \textit{italics}). Best scores for each cohort are reported in \textbf{bold} font. Dashes (\hyp{}) denote undefined result because of single-class cohorts --- TCGA\hyp{}LGG, TCGA\hyp{}GBM, and UPenn\hyp{}GBM. \(^\ast\)denote F1-macro and \(^\dagger\)denote F1-binary.}
  \label{tab:metrics_two_models}
  \setlength{\tabcolsep}{1.6pt}
  \renewcommand{\arraystretch}{1.28}
  \begingroup\fontsize{7.5}{9}\selectfont
  \begin{tabularx}{\textwidth}{@{}l *{14}{Y}@{}}
  \toprule
  \textbf{Dataset} &
  \multicolumn{2}{c}{\textbf{UCSF\textendash PDGM}} &
  \multicolumn{2}{c}{\textbf{TCGA\textendash LGG}} &
  \multicolumn{2}{c}{\textbf{TCGA\textendash GBM}} &
  \multicolumn{2}{c}{\textbf{EGD}} &
  \multicolumn{2}{c}{\textbf{IvyGAP}} &
  \multicolumn{2}{c}{\textbf{UPenn\textendash GBM}} &
  \multicolumn{2}{c}{\textbf{Overall}} \\
  \cmidrule(lr){2-3}\cmidrule(lr){4-5}\cmidrule(lr){6-7}\cmidrule(lr){8-9}\cmidrule(lr){10-11}\cmidrule(lr){12-13}\cmidrule(lr){14-15}
  \textbf{Model} & GPT\mbox{-}4o & GPT\mbox{-}5 & GPT\mbox{-}4o & GPT\mbox{-}5 & GPT\mbox{-}4o & GPT\mbox{-}5 & GPT\mbox{-}4o & GPT\mbox{-}5 & GPT\mbox{-}4o & GPT\mbox{-}5 & GPT\mbox{-}4o & GPT\mbox{-}5 & GPT\mbox{-}4o & GPT\mbox{-}5 \\
  \midrule
  \multirow{2}{*}{\textbf{Accuracy}}
   & 94.13 & \textbf{94.36} & 61.90 & \textbf{71.43} & \textbf{94.34} & 92.45 & 89.72 & \textbf{91.52} & 93.55 & 93.55 & \textbf{97.33} & 94.65 & 91.87 & \textbf{92.15} \\
   & \tiny\textit{(91.54--95.96)} & \tiny\textit{(91.80--96.15)} &
     \tiny\textit{(51.22--71.55)} & \tiny\textit{(61.00--79.98)} &
     \tiny\textit{(88.20--97.38)} & \tiny\textit{(85.81--96.13)} &
     \tiny\textit{(86.30--92.36)} & \tiny\textit{(88.33--93.90)} &
     \tiny\textit{(79.28--98.21)} & \tiny\textit{(79.28--98.21)} &
     \tiny\textit{(95.15--98.54)} & \tiny\textit{(91.88--96.51)} &
     \tiny\textit{(90.34--93.18)} & \tiny\textit{(90.64--93.44)} \\
  \midrule
  \multirow{2}{*}{\textbf{Sensitivity}}
   & 78.67 & \textbf{82.67} & 61.90 & \textbf{71.43} & --- & --- & 74.81 & \textbf{86.67} & 0.00 & 0.00 & --- & --- & 71.86 & \textbf{81.02} \\
   & \tiny\textit{(68.12--86.42)} & \tiny\textit{(72.57--89.58)} &
     \tiny\textit{(51.22--71.55)} & \tiny\textit{(61.00--79.98)} &
     \tiny\textit{---} & \tiny\textit{---} &
     \tiny\textit{(66.88--81.38)} & \tiny\textit{(79.91--91.40)} &
     \tiny\textit{(0.00--79.35)} & \tiny\textit{(0.00--79.35)} &
     \tiny\textit{---} & \tiny\textit{---} &
     \tiny\textit{(66.48--76.69)} & \tiny\textit{(76.15--85.08)} \\
  \midrule
  \multirow{2}{*}{\textbf{Specificity}}
   & \textbf{97.28} & 96.74 & --- & --- & \textbf{94.34} & 92.45 & \textbf{97.63} & 94.94 & 96.67 & 96.67 & \textbf{97.33} & 94.65 & \textbf{97.08} & 95.05 \\
   & \tiny\textit{(95.07--98.52)} & \tiny\textit{(94.39--98.12)} &
     \tiny\textit{---} & \tiny\textit{---} &
     \tiny\textit{(88.20--97.38)} & \tiny\textit{(85.81--96.13)} &
     \tiny\textit{(94.94--98.91)} & \tiny\textit{(90.49--96.39)} &
     \tiny\textit{(83.33--99.41)} & \tiny\textit{(83.33--99.41)} &
     \tiny\textit{(95.15--98.54)} & \tiny\textit{(91.88--96.51)} &
     \tiny\textit{(95.93--97.91)} & \tiny\textit{(93.63--96.17)} \\
  \midrule
  \textbf{F1 Score} &
  0.89$^\ast$ & \textbf{0.90}$^\ast$ & 0.76$^\dagger$ & \textbf{0.83}$^\dagger$ & \textbf{0.97}$^\dagger$ & 0.96$^\dagger$ & 0.88$^\ast$ & \textbf{0.91}$^\ast$ & 0.97$^\dagger$ & 0.97$^\dagger$ & \textbf{0.99}$^\dagger$ & 0.97$^\dagger$ & 0.87$^\ast$ & \textbf{0.88}$^\ast$ \\
  \bottomrule
  \end{tabularx}
  \endgroup
\end{table*}

\subsection{Datasets}
We assembled a multi-cohort dataset (\(N=1427\)) from six public repositories (Table \ref{tab:overall_characteristics}): UCSF-PDGM (\(n=443\), \(83\)\% IDH-wildtype, GBM) \cite{calabrese2022university}, TCGA-LGG (\(n=84\), \(100\)\% IDH-mutant, LGGs) \cite{bakas2017segmentation}, TCGA-GBM (\(n=106\), \(100\)\% IDH-wildtype, GBM) \cite{scarpace2016cancer}, EGD (\(389\), \(35\)\% IDH-mutant, LGGs and HGGs) \cite{van2021erasmus}, Ivy GAP (\(n=31\), \(97\)\% IDH-wildtype, GBMs) \cite{shah2016data}, and UPenn-GBM (\(n=374\), \(100\)\% IDH-wildtype, GBM) \cite{bakas2021multi}. Ground-truth multi-class segmentation maps were available for 
\(66\)\% subjects. Segmentations labels include peritumoral edema (ED), enhancing core (ET), and non-enhancing core (NET). Clinical and genomic information were retrieved from respective repositories. Inclusion criteria included availability of IDH mutation status and preoperative 3D multiparametric MRI scans (FLAIR, T1-weighted, T1-CE, and T2-weighted).

\subsection{Statistical Analysis}
We reported accuracy, sensitivity, specificity, and F1 score for predictive performance. \(95\)\% Wilson confidence intervals were also reported. To assess generalizability, F1-macro was used for cohorts with mixed IDH genotypes, and F1-binary for cohorts with a single genotype. In the ablation study, Recall evaluated IDH genotype prediction per molecular subtype, while the geometric mean of Recall measured balanced performance across subtypes. 

\subsection{System and LLM Specifications}
\label{sec:implementation details}
All experiments were performed on a workstation equipped with a NVIDIA GeForce GTX \(1080\) Ti (\(12\) GB VRAM) and a \(64\) GB CPU RAM. End-to-end processing for each subject (I/O, preprocessing, and feature extraction in JSON) took 5-7 minutes of wall clock time. We queried GPT-5 Chat (\texttt{gpt-5-chat-latest}) and GPT-4o (\texttt{gpt-4o}) via the Open AI API (temperature = \(0.0\), max tokens = \(1028\)). Each subject was processed independently, with the client re-initialized for every case to prevent information leakage. Mean latency (per-subject) was 5.5s for GPT-5 and 6s for GPT-4o, with corresponding average API costs of 0.002 USD (GPT-5) and 0.004 USD (GPT-4o).

\subsection{Results}
Table~\ref{tab:metrics_two_models} summarizes performance comparison of GPT-4o and GPT-5 models across multiple datasets. Across all datasets, GPT-5 generally outperformed GPT-4o in classification performance metrics. Averaged across cohorts, GPT-5 achieved a slightly higher overall accuracy (\(92.15\)\%) compared to GPT-4o (\(91.87\)\%) and demonstrated improved sensitivity (\(81.02\)\% vs. \(71.86\)\%). Specificity remained comparable between models (\(95.05\)\% for GPT-5 vs. \(97.08\)\% for GPT-4o). At the cohort level, GPT-5 achieved the highest accuracy for UCSF–PDGM (\(94.36\)\%), TCGA–LGG (\(71.43\)\%), and EGD (\(91.52\)\%), while GPT-4o slightly outperformed in TCGA–GBM (\(94.34\)\%) and UPenn–GBM (\(97.33\)\%). F1 scores followed similar trends, with GPT-5 yielding marginal improvements across most datasets (overall F1-macro = \(0.88\) vs. \(0.87\) for GPT-4o). These results indicate that GPT-5 maintains strong generalization across heterogeneous cohorts and provides consistent improvements in sensitivity and F1 performance, suggesting enhanced capability for balanced classification.
Strong predictive performance was observed in datasets with available ground-truth multi-class segmentation maps (UCSF–PDGM: F1 = \(0.90\); UPenn–GBM: F1 = \(0.97\)), as well as in datasets lacking ground-truth segmentation maps (EGD: F1 = \(0.91\)). 

\subsection{Ablation Study}
The ablation study, presented in Table~\ref{tab:gpt5_ablation}, evaluated the impact of removing specific group of features, from the JSON file, on IDH prediction. The study was conducted using GPT-5, which was both the best-performing and more economical model compared to GPT-4o. Recall was assessed separately for IDH-mutant astrocytoma (Astro) and oligodendroglioma (Oligo), since their imaging phenotypes differ despite a shared IDH mutation status \cite{jain2020real}. A combined metric could mask subtype-specific effects. 

The baseline model achieved strong performance across Astro, Oligo, and GBM, with an overall geometric mean recall of \(0.83\). Removing volumetric measures significantly decreased predictive performance (overall recall \(0.68\)), indicating that these features are critical for balanced predictions across molecular subtypes. Excluding location features noticeably reduced Oligo recall (\(0.77\) vs. \(0.80\)). Removing mass effect attributes slightly lowered overall performance compared to the baseline (\(0.81\) vs. \(0.83\)). Omitting T2–FLAIR mismatch features had a significant negative impact on GBM prediction (recall \(0.69\) vs. \(0.87\)) but improved predictive performance for Oligo (\(0.92\)) and Astro (\(0.92\)), highlighting the poor sensitivity of the T2–FLAIR mismatch signature for IDH prediction, across diverse cohorts. Removing morphological features substantially reduced Oligo recall (\(0.69\) vs. \(0.80\)) while improving GBM recall (\(0.95\) vs. \(0.87\)). Finally, augmenting the baseline model with clinical data improved GBM recall to \(0.95\) and increased the overall recall to \(0.84\), suggesting that incorporating clinical information enhances model robustness and accuracy across subtypes.

\newcolumntype{Y}{>{\centering\arraybackslash}X}
\begin{table}[t]
\centering
\footnotesize
\setlength{\tabcolsep}{4pt}
\renewcommand{\arraystretch}{1.15}
\caption{GPT-5 ablation study results}
\label{tab:gpt5_ablation}
\begin{tabularx}{0.95\linewidth}{@{}lYYYY@{}}
\toprule
& \textbf{Astro Recall} & \textbf{Oligo Recall} & \textbf{GBM Recall} & \textbf{Geometric Mean} \\
\midrule
Baseline                 & 0.84 & 0.80 & 0.87 & 0.83 \\
\addlinespace[2pt]
\hspace{0.5em}-- Volumetric Measures   & 0.83 & 0.85 & 0.44 & 0.68 \\
\hspace{0.5em}-- Location Features     & 0.82 & 0.77 & 0.82 & 0.80 \\
\hspace{0.5em}-- Mass Effect           & 0.81 & 0.75 & 0.87 & 0.81 \\
\hspace{0.5em}-- T2--FLAIR Mismatch    & 0.92 & 0.92 & 0.69 & 0.84 \\
\hspace{0.5em}-- Tumor Morphology      & 0.81 & 0.69 & 0.96 & 0.81 \\
\midrule
Baseline + Clinical      & 0.85 & 0.74 & 0.95 & 0.84 \\
\bottomrule
\end{tabularx}
\end{table}

\section{Discussion}
This work demonstrates that LLMs can perform zero-shot IDH genotype classification of brain gliomas from structured, imaging-derived semantic attributes and quantitative features. Both GPT-4o and GPT-5 achieved high accuracy across six public cohorts without fine-tuning, indicating strong generalization to heterogeneous data. GPT-5 provided higher sensitivity and balanced F1 scores, particularly for IDH-mutant tumors, suggesting improved integration of imaging and contextual cues rather than relying on isolated/simplistic radiologic features. Moreover, strong predictive performance was also observed in subjects without ground-truth multiclass segmentation masks (e.g., EGD), demonstrating the model’s ability to accurately predict IDH genotype even in the absence of complete annotations.

Out of \(1,427\) total subjects, GPT-4o and GPT-5 agreed on \(1,349\) cases (\(1,274\) correct and \(75\) incorrect predictions) and disagreed on \(78\) cases. Inspection of the \(78\) discordant cases revealed notable patterns: GPT-4o performed better in high-grade, aggressive gliomas characterized by distinct imaging features, deep gray matter involvement, and extensive edema, while GPT-5 excelled in low-grade gliomas with complex or non-enhancing lesions. Overall, GPT-5 demonstrated superior predictive performance across molecular subtypes, suggesting it should be the preferred model when context-driven phenotype interpretation is required.

Ablation results revealed that volumetric measures were the most critical features for accurate IDH genotype prediction, while T2–FLAIR mismatch and morphology contributed subtype-specific value. The modest improvement after incorporating age and gender (clinical metrics) reaffirmed that simple clinical context can enhance model stability and recall. These findings suggest that LLMs can synthesize structured imaging and clinical information for enhanced prediction of IDH genotype in brain gliomas. 

Our work has several limitations: (i) The study focused on zero-shot LLM-based inference without fine-tuning, which may have limited optimal model performance. 
(ii) The zero-shot setup also limited interpretability beyond the prompt-defined reasoning framework. 
(iii) While we observed improved performance on datasets without manual annotations, a more extensive analysis is needed to assess the impact of different automated segmentation algorithms. (iv) Moreover, the robustness of genotype prediction to feature extraction algorithms is also required. (v) We evaluated only proprietary, API-accessible LLMs; behavior may differ for open-weight models.
Future work will (i) co-design the attribute set with neuro-oncology experts to enhance clinical validity and (ii) benchmark the pipeline on high-performing open-source medical LLMs to assess reproducibility as well as cost and accessibility considerations.

\section{Conclusion}

In summary, this study demonstrated that LLMs, especially GPT-5, can accurately predict IDH genotype in brain gliomas from structured imaging-derived features and basic clinical vairables without fine-tuning, generalizing across diverse cohorts. GPT-5 outperformed GPT-4o, especially for LGGs or complex lesions; volumetrics were the strongest predictors, supplemented by subtype-specific imaging markers and clinical information. These findings highlight the potential of foundation models to integrate multimodal information for zero-shot, non-invasive genotype classification. 


\section{Compliance with Ethical Standards}
We conducted a retrospective study using only publicly available datasets. All multimodal datasets were fully de-identified, and the necessary Institutional Review Board (IRB) approvals were obtained by the respective centers that made the datasets publicly accessible. This study was carried out in accordance with the TRIPOD-LLM guidelines.

\section{Acknowledgments}
The authors gratefully acknowledge \href{https://talhaahmed2000.github.io}{Talha Ahmed} (Algorithms in Theory and Practice Lab) for sustained support and valuable discussions. This work was supported by a grant from the Higher Education Commission of Pakistan as part of the National Center for Big Data and Cloud Computing. Email: hassan.mohyuddin@lums.edu.pk

\bibliographystyle{IEEEbib}
\bibliography{main}

\end{document}